\documentclass[11pt]{article}
\usepackage[dvips]{color}
\usepackage{epsfig}
\usepackage{amsmath}
\usepackage{graphicx}
\date{}

\begin{document}

\title{{\bf Affine Quantization of the Interior Schwarzschild Black Hole}}

\author{Morteza Bajand\thanks{%
e-mail: m.bajand@iau.ac.ir}\,\, and Babak Vakili\thanks{email:
ba.vakili@iau.ac.ir (corresponding author)}\\\\{\small {\it
Department of Physics, CT.C., Islamic Azad
University, Tehran, Iran}}} \maketitle

\begin{abstract}
In this paper, we investigate the Hamiltonian formulation of a spherically symmetric spacetime that corresponds to the interior of a Schwarzschild black hole. The resulting phase space involves two independent dynamical variables along with their conjugate momenta. We quantize the associated minisuperspace using the affine quantization method, which is particularly suited for systems with positive-definite configuration variables. This approach leads to a Wheeler–DeWitt (WDW) equation that governs the affine quantum dynamics of the black hole interior. Analytical and approximate solutions to the WDW equation are obtained, allowing us to construct the corresponding wave function. We then explore whether the quantum effects encoded in this wave function can lead to the avoidance of classical singularities.
\vspace{5mm}\noindent\vspace{0.8mm}\newline Keywords:
Interior Schwarzschild black hole; Affine quantization; Quantum black hole
\end{abstract}

\section{Introduction}
Black hole physics plays a central role in fundamental theoretical physics, as it is associated with some of the most profound and unresolved problems in modern science \cite{1}. Among these, the presence of singularities inside event horizons and the black hole information paradox are particularly prominent \cite{2,3}. Various approaches have been proposed to address these issues, one of which involves the quantization of gravitational models and treating black holes as quantum systems.
In this context, black hole thermodynamics—arising from the application of quantum field theory to curved spacetime—provides a compelling framework. According to this view, Hawking radiation emerges from quantum fluctuations near the event horizon~\cite{4}. These insights naturally support the idea that black holes possess a quantum mechanical nature, and like any other quantum system, their physical state should be described by a wave function\cite{5,6}.
Given the conceptual importance of wave functions in quantum gravity, studying them provides a meaningful testbed for exploring different approaches to quantum gravity. Nevertheless, despite significant advances in quantum mechanics, a complete and consistent quantum theory of gravity remains an open challenge.

On the other hand, the interior of the Schwarzschild black hole has attracted considerable interest for both mathematical and physical reasons \cite{7}. Several authors have suggested that the black hole interior can be regarded as a cosmological solution to the Einstein field equations and have explored its implications in the context of cosmology \cite{Bre}. One particularly intriguing consequence of this perspective is that techniques developed for quantizing cosmological models can also be applied to the spacetime inside a black hole~\cite{Boss}.
Mathematically, the interior metric can be derived from the exterior Schwarzschild solution via a coordinate transformation. A notable feature of general relativity is its ability to generate dynamical (cosmological) solutions from known static ones. This transformation is often based on a diffeomorphism between static and cosmological spacetimes, as discussed in \cite{Que}. Specifically, exchanging the roles of the temporal and radial coordinates, \( t \leftrightarrow r \), converts the Schwarzschild metric into a time-dependent form suitable for describing the black hole interior. At the event horizon, the light cone structure changes such that for \( r < r_s \), the radial coordinate \( r \) becomes timelike and \( t \) becomes spacelike. As a result, the interior geometry evolves dynamically, and this correspondence enables the formulation of a quantum theory of black holes based on techniques from quantum cosmology.

It is worth noting that the interior of the Schwarzschild black hole has also been studied extensively in the context of Loop Quantum Gravity (LQG). In particular, several works have employed the isometry between the Schwarzschild interior and the Kantowski–Sachs cosmological model to implement loop quantization techniques and analyze the fate of the classical singularity. See, for instance, \cite{ashtekar2005schwarzschild}-\cite{corichi2016loop}. While the quantization approach adopted in this work is based on affine quantization, rather than polymer quantization used in LQG, both frameworks aim at addressing the same fundamental issues concerning singularity resolution in quantum gravity.

In this paper, we investigate the Hamiltonian formulation of a time-dependent spherically symmetric spacetime whose dynamics correspond to the interior geometry of a Schwarzschild black hole. In this model, the metric functions are treated as independent dynamical variables, and together with their conjugate momenta, they form the corresponding phase space.
To quantize the system, we employ the affine quantization method, which is particularly well-suited for configuration spaces with positive-definite variables. Using the covariant integral approach, we derive the associated Wheeler–DeWitt (WDW) equation governing the quantum dynamics. We then formulate a Hamiltonian quantum theory by promoting the classical phase space variables to affine quantum operators. We provide analytical solutions to the WDW equation and obtain explicit expressions for the wave function of the quantum black hole. Our analysis demonstrates that the singularity present in the classical model is effectively resolved through the quantum dynamics encoded in the affine quantization scheme.

\section{The model}
In this section, we briefly describe the gravitational model under consideration, beginning with the Einstein–Hilbert action

\begin{equation}\label{A}
S=\frac{1}{16 \pi G} \int d^4x \sqrt{-g}R+S_{YGB},
\end{equation}
where \( R \) is the Ricci scalar, \( g \) is the determinant of the metric tensor, and \( S_{YGB} \) denotes the York–Gibbons–Hawking boundary term. We consider a time-dependent spherically symmetric spacetime geometry described by the following metric

\begin{equation}\label{B}
ds^{2}=-\frac{N^{2}(t)}{\nu(t)} dt^{2}+\nu(t) dr^{2}+h^{2}(t)\left(d \vartheta^{2} + \sin^{2}\vartheta d\varphi^{2}\right).
\end{equation}
Here, \( N(t) \) is the lapse function, while \( \nu(t) \) and \( h(t) \) are functions of time that serve as the dynamical variables of the corresponding phase space. As is well known, inside a Schwarzschild black hole, the radial coordinate becomes timelike and the time coordinate becomes spacelike. As a result, the metric acquires time-dependent coefficients. This spacetime corresponds to the interior region of the Schwarzschild solution under the following identification

\begin{equation}\label{C}
N=1,\qquad   h(t)=t,\qquad  \nu(t)=\frac{2GM}{t}-1.  
\end{equation}    
We now turn to the Hamiltonian structure of the model. Substituting equation (\ref{B}) into (\ref{A}) and integrating over the spatial coordinates yields an effective Lagrangian in the minisuperspace parametrized by \( (\nu, h) \):

\begin {equation}\label{D}
L=- \frac{V_{0}}{8 \pi G} \left[\frac{1}{N}\left(h\dot{h}\dot{\nu}+\dot{h}^{2}\nu\right)-N\right],
\end{equation}
where \( V_0 \) denotes the spatial volume factor appearing in the action, which is treated as a finite constant. The equations of motion for \( h \) and \( \nu \) can be derived by varying the effective Lagrangian. However, the resulting Hamiltonian constraint does not take a suitable form for constructing the WDW equation corresponding to the quantum model \cite{6}. To recast the Lagrangian (\ref{D}) into a more tractable form, we introduce the following change of variables

\begin{equation}\label{E}
x-y=h,\qquad x+y=h\nu,
\end{equation}                 
according to which, Lagrangian (\ref{D}) takes the form 
\begin{equation}
L=-\frac{V_{0}}{8\pi G}\left[\frac{1}{N}\left(\dot{x}^{2}- \dot{y}^{2}\right)-N\right].
\end{equation}
Now, consider the following coordinate transformation \cite{Bre}

\begin{eqnarray}\label{F}
\left\{
\begin{array}{ll}
(x-y)^{\frac{1}{2}}=\frac{1}{2} (u-v),\\\\
(x+y)^{\frac{1}{2}}= \frac{1}{2} (u+v),
\end{array}
\right.
\end{eqnarray}
and also the lapse rescaling

\begin{equation}\label{G}
N=(x^{2}-y^{2})^{\frac{1}{2}} \widetilde{N}.
\end{equation}
by which the Lagrangian becomes

\begin{equation}\label{H}
L=-M_{Pl}^{2} V_{0}\widetilde{N}^{-1} \left(\dot{u}^{2} - \dot{v}^{2}\right) +M_{Pl}^{2} V_{0} \widetilde{N}\left(u^{2}-v^{2}\right).
\end{equation}
The momenta conjugate to the dynamical variables $u$ and $v$ are given by

\begin{equation}\label{K}
\Pi_{u}=-\frac{2 V_{0} M_{PL}^{2} }{\widetilde{N}} \dot{u},\hspace{5mm}\Pi_{v}= \frac{2 V_{0} M_{PL}^{2} }{\widetilde{N}} \dot{v},
\end{equation}
leading to the following Hamiltonian

\begin{equation}\label{I}
H=-\frac{\widetilde{N}}{4M_{Pl}^{2} V_{0}}\left(\Pi_{u}^{2} - \Pi_{v}^{2}\right)-V_{0} M_{Pl}^{2} \widetilde{N}\left(u^{2} - v^{2}\right),
\end{equation}
Also, the primary constraints is given by
 
\begin{equation}\label{J}
\Pi_{\widetilde{N}}=\frac{\partial L}{\partial  \widetilde{N}}=0.
\end{equation}The requirement that the primary constraints
should hold during the evolution of the system means that $\left\{\Pi_N,H\right\}=0,$ which leads to the secondary constraint

\begin{equation}\label{L}
{\cal H}=-\frac{1}{4M_{Pl}^{2} V_{0}}\left(\Pi_{u}^{2} - \Pi_{v}^{2}\right)- V_{0}M_{Pl}^{2}\left(u^{2} - v^{2}\right)=0.
\end{equation}
At this stage, the phase space structure, as well as the corresponding Lagrangian and Hamiltonian formulations, are fully specified. The classical dynamics of the system is then determined by the Hamiltonian equations of motion. As shown in \cite{6}, under the gauge fixing condition \( \tilde{N} = M_{\text{Pl}} \) and time reparameterization \( N dt = d\tau \), where \( N = M_{\text{Pl}}(u^2 - v^2)/4 \), the resulting classical equations reproduce the interior Schwarzschild black hole metric

\begin{equation}\label{M}
ds^2=-\left(\frac{2GM}{\tau}-1\right)^{-1}d\tau^2+\left(\frac{2GM}{\tau}-1\right)dr^2+\tau^2\left(d\vartheta^2+\sin^2\vartheta d\varphi^2\right).
\end{equation}

\section{A brief review of affine quantization}
In the standard canonical quantization framework, each pair of classical variables \( (q, p) \) in the phase space \( \textsf{R}^2 \) is promoted to a pair of Hermitian, non-commuting operators. This implies that the classical variables are assumed to range over the entire real line, \( (-\infty, +\infty) \). Accordingly, each pair of classical dynamical variables is replaced by a pair of self-adjoint operators \( (\hat{Q}, \hat{P}) \), acting on the Hilbert space of square-integrable complex-valued wave functions \( \Psi(q) \), with the following representation:
\begin{equation}\label{N}
[\hat{Q}, \hat{P}] = i\hbar, \qquad \hat{Q}\Psi(q) = q\Psi(q), \qquad \hat{P}\Psi(q) = -i\hbar \frac{\partial}{\partial q} \Psi(q).
\end{equation}
For any classical observable \( f(q,p) \), its quantum counterpart is formally given by \( f(\hat{Q}, \hat{P}) \), although operator ordering ambiguities generally arise and must be addressed.

However, there are important cases where the geometry of the phase space is different from \( \textsf{R}^2 \). In particular, in many gravitational systems such as black holes and cosmological models, the relevant phase space may be constrained—for instance, restricted to the upper half-plane

\begin{equation}\label{O}
\Pi_{+}=\textsf{R}_{+} \times \textsf{R}=\left\{(q,p)|q>0,p\in \textsf{R}\right\}.\end{equation}
In such scenarios, canonical quantization may not be well-suited, and alternative quantization methods must be considered. Since quantization methods rely on the underlying symmetry of the canonical pair \( (q, p) \), the standard canonical quantization based on the relations in equation (\ref{N}) is well suited for phase spaces of the form \( \{(q, p)\,|\, q, p \in \textsf{R} \} \). However, it is not appropriate for constrained phase spaces such as the one defined in equation (\ref{O}) \cite{Tw, Zach}. In such cases, canonical quantization often requires imposing additional conditions on the state vectors or modifying the Hamiltonian with extra terms in order to recover physically consistent results. A typical example is the quantum description of a particle in a box, which requires specific boundary conditions not naturally accounted for in standard quantization procedures.
One widely studied approach to address these limitations is affine quantization. This method has proven effective in the treatment of systems with strictly positive dynamical variables, including certain cosmological models \cite{Ber} and black hole configurations \cite{BH}. In the affine covariant integral quantization approach, one begins by considering the affine group \cite{Alm}, which forms the mathematical basis of this quantization scheme

\begin{equation}\label{P}
\Pi_{+}=\textsf{R}_{+} \times \textsf{R}=\left\{(q,p)|q>0,p\in \textsf{R}\right\},\hspace{5mm}(q,p)(q',p'):=\left(qq',\frac{p'}{q}+p\right),
\end{equation}with unity $(1,0)$ and inverse $(q,p)^{-1}=(1/q,-qp)$. The affine group possesses a rich algebraic structure that makes it well suited for the quantization of phase spaces with positive-definite variables. For a detailed account of the mathematical and physical properties of this group, see \cite{Com}. Within this quantization framework, a classical observable \( f(q, p) \) is mapped to a quantum operator acting on the Hilbert space of square-integrable functions defined on the half-plane. To formalize this, we introduce the Hilbert space \( H_\gamma \), defined as follows

\begin{equation}\label{Q}
H_{\gamma}=\textsf{L}^2 \left(\textsf{R}_{+},\frac{dx}{x^{\gamma +1}}\right).
\end{equation}Equipping this Hilbert space with the measure \( d\mu = dx / x^{\gamma+1} \), one finds that its basis functions satisfy both the orthogonality and completeness relations

\begin{equation}\label{R}
\langle x |x' \rangle =x^{\gamma+1}\delta (x-x'),\hspace{5mm} \int_{0}^{\infty}\frac{dx}{x^{\gamma+1}} | x \rangle \langle x' |=I.
\end{equation}To construct the operator representation in this Hilbert space, we make use of the coherent state formalism, as described in \cite{Gaz}

\begin{equation}\label{S}
|q,p \rangle=U(q,p) |\psi \rangle,
\end{equation}
where $U(q,p)$, is an unitary irreducible representation of affine group and $|\psi \rangle \in H_{\gamma} \cap H_{\gamma+1}$, is a fiducial vector (wavelets) such that 

\begin{equation}\label{T}
\left(U(q,p)\psi\right)(x)=e^{ipx} q^{\frac{\gamma}{2}} \psi\left(\frac{x}{q}\right)\Rightarrow \langle x | q,p \rangle =e^{ipx} q^{\frac{\gamma}{2}} \psi\left(\frac{x}{q}\right).
\end{equation}
Within the framework of affine quantization, each classical observable \( f(q, p) \) is associated with a unique quantum operator \( A_f \), defined as

\begin{equation}\label{U}
A_{f}=\int_{0}^{\infty}  \int_{- \infty}^{\infty}  \frac{dpdq}{2 \pi c_{\gamma}^{n}} f(q,p) |q,p\rangle \langle q,p|,
\end{equation}
where $c_{\gamma}^{n}$ is defined as

\begin{equation}\label{U1}
c_{\gamma}^{n} (\psi)= \int \frac{dx}{x^{\gamma +2}} |\psi^{(n)} (x)|^{2},
\end{equation}
and $\psi^{(n)}$ is the $n^{th}$ derivative of $\psi$. In order to assign explicit expressions to the canonical operators \( \hat{Q} \), \( \hat{P} \), and their powers, we consider functions $\phi(x): \textsf{R}_{+}\rightarrow \textsf{R}$, and study the action of the operator \( A_f \) on the state \( |\phi\rangle \) in the position basis \( |x\rangle \), given by

\begin{equation}\label{V}
\langle x|A_f|\phi \rangle=\int_{0}^{\infty}  \int_{- \infty}^{\infty}  \frac{dqdp}{2 \pi c_{\gamma}^{n}} f(q,p)\int_{0}^{\infty}\frac{dx'}{x'^{\gamma+1}}q^{\gamma}e^{ip(x-x')}\psi \left(\frac{x}{q}\right)\psi \left(\frac{x'}{q}\right)\phi(x').
\end{equation}With the use of this relation we can get the affine quantization counterparts of the canonical operators (\ref{N}) as 

\begin{equation}\label{W}
A_{q^{\beta}}=\frac{c_{\gamma + \beta}}{c_{\gamma}} \hat{Q}^{\beta},
\end{equation}

\begin{equation}\label{X}
A_{p}= \hat{P}+i\left(\frac{\gamma +1}{2}\right) \hat{Q}^{-1},
\end{equation}

\begin{equation}\label{Y}
A_{p^{2}}=\hat{P}^{2} +i\left(\gamma +1\right) \hat{Q}^{-1} \hat{P} +\left[\frac{c_{\gamma -2}^{(1)}}{c_{\gamma}} - \frac{(\gamma +1)(\gamma+2)}{2}\right]\hat{ Q}^{-2},
\end{equation}
which are a direct result of the evaluation of $\langle x|A_{q^{\beta}}|\phi \rangle$,  $\langle x|A_{p}|\phi \rangle$ and  $\langle x|A_{p^2}|\phi \rangle$. It can be seen that the operator corresponding to the classical monomial \( q^\beta \), namely \( A_{q^\beta} \), is proportional to the canonical operator \( \hat{Q}^\beta \), differing only by a constant factor. In contrast, the operators associated with momentum and its powers exhibit significant deviations from their canonical counterparts. Finally, using equation (\ref{U}), one can derive an expression for computing the expectation value of the operator \( A_f \) with respect to the coherent state \( |q, p\rangle \), given by

\begin{equation}\label{Z}
\langle A_f \rangle=\langle q,p |A_f|q,p \rangle,
\end{equation}which somehow, may be considered as the quantum corrections to the classical observable $f(q,p)$. A straightforward calculation based on the relations (\ref{S}), (\ref{T}) and (\ref{U}) yields 

 \begin{eqnarray}\label{AA}
\langle A_f \rangle&=&\int_{0}^{\infty}  \int_{- \infty}^{\infty}  \frac{q^{\gamma}q'^{\gamma}}{2 \pi c_{\gamma}^{n}}dq'dp'\int_{0}^{\infty}\int_{0}^{\infty}\frac{dx dx'}{x^{\gamma+1}x'^{\gamma+1}}f(q',p')\\ \nonumber &\times& e^{ip(x-x')}e^{-ip'(x-x')}\psi\left(\frac{x}{q}\right)\psi\left(\frac{x'}{q}\right)\psi\left(\frac{x}{q'}\right)\psi\left(\frac{x'}{q'}\right).
\end{eqnarray}
The framework for the application of affine quantization has now been established. In the following sections, we apply this quantization scheme to the gravitational model introduced in Section 2, in order to investigate how the classical description is modified by quantum effects.

\section{Affine quantization of the model}
In this section, we implement the affine quantization procedure to analyze the quantum interior of the black hole as described in the preceding sections. To this end, we quantize the dynamical variables of the model using the WDW equation, given by ${\cal H}\Psi(u,v)=0$, where \( \cal{H} \) is the operator form of the Hamiltonian introduced in equation (\ref{L}), and \( \Psi(u,v) \) is the wave function of the black hole, expressed in terms of the variables \( u \) and \( v \) as

\begin{eqnarray}\label{AB}
\left\{
\begin{array}{ll}
u=\sqrt{h \nu}+\sqrt{h},\\\\
v=\sqrt{h \nu}-\sqrt{h}.
\end{array}
\right.
\end{eqnarray}
Since $u \in \textsf{R}_{+}$ and $v \in \textsf{R}$, we apply affine quantization to the variable \( u \) and use the standard canonical quantization for \( v \). Accordingly, using the relations (\ref{W})–(\ref{Z}), we obtain

\begin{eqnarray}\label{AC}
\left\{
\begin{array}{ll}
A_{\Pi_{u}^{2}}= \Pi_{u}^{2} -i u^{-1} \Pi_{u} +(\frac{c_{-4}}{c_{-2}}) u^{-2},\\\\
A_{u^{2}}=\frac{c_{0}}{c_{-2}} u^{2},
\end{array}
\right.
\end{eqnarray}
in which we have chosen $\gamma=-2$. Under these conditions the WDW equation takes the form

\begin{equation}\label{AD}
\left[-\left(A_{\Pi_{u}^{2}} + A_{u}^{2}\right)+ \left(\Pi_{v}^{2} + v^{2}\right)\right]\Psi(u,v)=0.
\end{equation}
With the replacement $\Pi_v \rightarrow -i\frac{\partial}{\partial v}$, and the use of (\ref{AC}), this equation reads

\begin{equation}\label{AE}
\left(\frac{\partial^{2}}{\partial u^{2}}+u^{-1}\frac{\partial}{\partial u}-\frac{c_{-4}}{c_{-2}} u^{-2}-\frac{c_{0}}{c_{-2}}u^2-\frac{\partial^2}{\partial v^{2}} +v^{2}\right)\Psi(u,v)=0.
\end{equation}
The solutions of the above differential equation are separable into the form $\Psi(u,v)=U(u)V(v)$, leading to

\begin{equation}\label{AF}
\left(\frac{\partial^{2}}{\partial u^{2}}+u^{-1}\frac{\partial}{\partial u}-\frac{c_{-4}}{c_{-2}}u^{-2}-\frac{c_{0}}{c_{-2}} u^2\right)U(u)=EU(u),
\end{equation}

\begin{equation}\label{AG}
\left(-\frac{\partial^{2}}{\partial v^{2}}+ v^{2}\right)V(v)=EV(v),
\end{equation}
where $E$ is a separation constants. Equation (\ref{AF}), after rearranging it in the form  

\begin{equation}\label{AH}
\left(u^{2} \frac{\partial^{2}}{\partial u^{2}} +u \frac{\partial}{\partial u}-A-B u^{4}-E u^{2}\right)U(u)=0,
\end{equation}
where $A=( \frac{c_{-4}}{c_{-2}})$ and $B=(\frac{c_{0}}{c_{-2}})$, admits the solutions

\begin{equation}\label{AI}
U(u)=2^{\frac{\beta}{2}}e^{-\frac{z}{2}} u^{\beta -1}\left[c_{1} {\cal L}_{-\alpha}^{\beta-1} (z)+ c_{2} {\cal U}(\alpha, \beta; z)\right],
\end{equation}
where ${\cal U}(\alpha, \beta; z)$ is the confluent Hypergeometric function and ${\cal L}_{\xi}^{\mu} (z)$ is the generalized Laguerre polynomial. Also, $\alpha$, $\beta$ and $z$ are defined as

\begin{equation}\label{AJ}
\alpha= \frac{2B +2 \sqrt{-A} B+ \sqrt{B} E}{4B},\hspace{5mm}\beta=1+ \sqrt{-A},\hspace{5mm}z= \sqrt{B} u^{2}.
\end{equation}
If we use the relation between the Laguerre and hypergeometric functions as  
$L_{\xi}^{\mu} (z)= \frac{(-1)^{\xi}}{\xi!} {\cal U}(\alpha, \beta; z)$, the solution for $U(u)$ takes the form

\begin{equation}\label{AK}
U(u)=2^{\frac{\beta}{2}}e^{- \frac{z}{2}} u^{\beta -1} \left[c_{1} \frac{(-1)^{\xi}}{\xi!} + c_{2}\right] {\cal U}(\alpha, \beta;z),
\end{equation}where $\xi=-\alpha$. The solution can be expressed in terms of the Kummer function \( M(\alpha, \beta; z) \), using the identity provided in \cite{Abra}

\begin{equation}\label{AL}
{\cal U}(\alpha, \beta;z)=\frac{\pi}{\sin \pi \beta}\left[\frac{M(\alpha,\beta;z)}{\Gamma(1+\alpha-\beta)\Gamma(\beta)}-z^{1-\beta}\frac{M(1+\alpha-\beta,2-\beta;z)}{\Gamma(\alpha)\Gamma(2- \beta)}\right],
\end{equation}
and the power expansion of the Kummer function as

\begin{equation}\label{AM}
M(\alpha,\beta;z)=1+ \frac{\alpha}{\beta}z+\frac{\alpha (\alpha +1)}{\beta (\beta +1) 2!}z^2+.....,
\end{equation}
which finally converts the expression (\ref{AK}) as below

\begin{equation}\label{AN}
U_{E}(u)=\frac{2^{\frac{\beta}{2}}e^{- \frac{z}{2}} u^{\beta-1}}{\sin \pi \beta}\pi \left[c_{1} \frac{(-1)^{\xi}}{\xi!} + c_{2}\right] \left[\frac{1+\frac{\alpha}{\beta}z}{\Gamma(1+\alpha - \beta) \Gamma(\beta)}-z^{1- \beta}\frac{1+\frac{(1+\alpha-\beta)}{2- \beta}z}{\Gamma(\alpha) \Gamma(2-\beta)}\right].
\end{equation}
Returning to equation (\ref{AG}), the solutions can be expressed in terms of the parabolic cylinder functions as follows

\begin{equation}\label{AO}
V(v)= d_{1}\cos \pi \left(\frac{1}{4}- \frac{1}{2} E\right)Y_{1}-d_{2}\sin \pi \left(\frac{1}{4}- \frac{1}{2}E\right)Y_{2},
\end{equation}
where $Y_{1}$ and $Y_{2}$ are auxiliary functions defined as

\begin{equation}\label{AP}
Y_{1}=\sqrt{\pi}\frac{\sec \pi (\frac{1}{4}-\frac{1}{2} E)}{2^{- \frac{1}{2}E+\frac{1}{4}} \Gamma(\frac{3}{4}-\frac{1}{2} E)}e^{- v^{2}}M\left(-\frac{1}{2} E+\frac{1}{4} , \frac{1}{2};2 v^{2}\right),
\end{equation}

\begin{equation}\label{AQ}
Y_{2}=\sqrt{\pi}\frac{\csc \pi (\frac{1}{4}-\frac{1}{2} E)}{2^{-\frac{1}{2} E-\frac{1}{4}} \Gamma(\frac{1}{4}-\frac{1}{2} E)}2v e^{- v^{2}} M\left(-\frac{1}{2} E+\frac{3}{4}, \frac{3}{2};2 v^{2}\right).
\end{equation}
Here, $d_{1}$ and $d_2$ are integration constants, which are fixed as $d_1=-d_2=1$, for simplicity. Again, considering the Kummer functions as

\begin{equation}\label{AR}
M\left(-\frac{1}{2} E+\frac{1}{4},\frac{1}{2};2 v^{2}\right)=1+\left(\frac {1}{2}-E\right)(2 v^{2}),
\end{equation}

\begin{equation}\label{AS}
M\left(-\frac{1}{2}E+\frac{3}{4},\frac{3}{2};2 v^{2}\right)=1+\frac{1}{3}\left(\frac{3}{2}-E\right)(2 v^{2}),
\end{equation}
we get the solution for $V(v)$ as 

\begin{equation}\label{AT}
V_{E}(v)=\frac{\sqrt{\pi} e^{- v^{2}} \left[1+(\frac{1}{2}-E) (2 v^{2})\right]}{2^{-\frac{1}{2} E+\frac{1}{4}}\Gamma(\frac{3}{4}-\frac{1}{2} E)}-\frac{2\sqrt{\pi}v e^{- v^{2}} \left[1+\frac{1}{3}(\frac{3}{2}-E)(2 v^{2})\right]}{2^{-\frac{1}{2}E-\frac{1}{4}}\Gamma(\frac{1}{4}-\frac{1}{2} E)}.
\end{equation}
To summarize, the eigenfunctions of the WDW equation can be written as

\begin{equation}\label{AU}
\Psi_{E}(u,v)=U_{E}(u)V_{E}(u,v),
\end{equation}where $U_{E}(u)$ and $V_{E}(v)$, are given by the relations (\ref{AN}) and (\ref{AT}). The general solution can then written as a suitable superposition of these eigenfunctions, which will be considered in the next section. 
 
\section{The wave function: approximately analytical solutions} 
\begin{figure}
\includegraphics[width=2.5in]{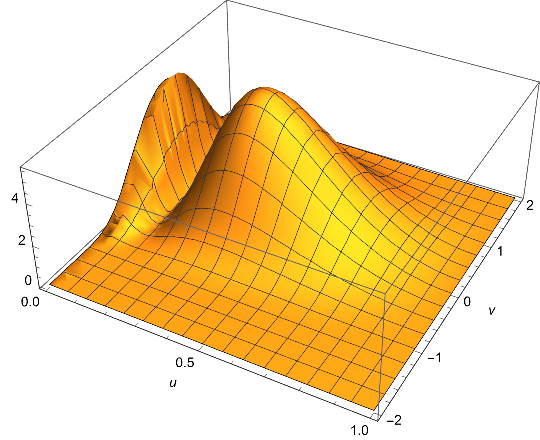}\hspace{25mm}\includegraphics[width=2in]{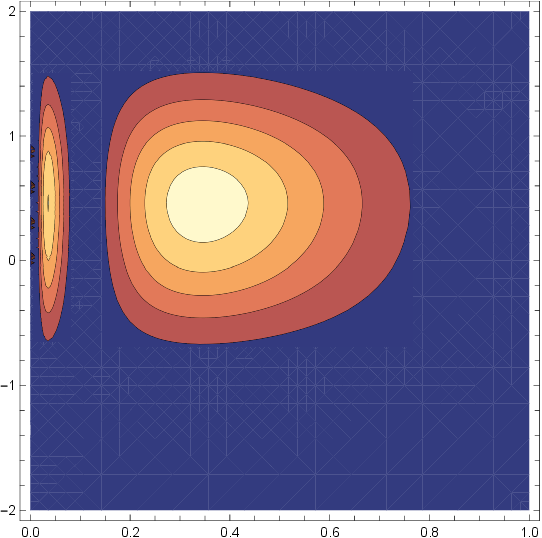}\\
\caption{The square of the wave function and its contour plot.}\label{fig1}
\end{figure}

In the previous section, we derived the eigenfunctions of the WDW equation. As noted earlier, the general solution is given by a superposition of these eigenfunctions, expressed as
 
\begin{equation}\label{AV}
\Psi (u,v)=U(u)V(v)=\int_{E}A_1(E)U_{E}(u)dE \int_{E}A_2(E)V_{E}(v)dE,
\end{equation}
where $ A_1(E)$ and $A_2(E)$ are appropriate weight functions used to construct the wave packets. To proceed, we apply a series of simplifying assumptions. First, we set $c_{1}=0$ and $c_{2}=1$, in the expression for $U_{E}(u)$. Next, we employ the asymptotic approximation $\Gamma(x+y) \approx \Gamma(x) x^{y}$, valid in the limit $x\rightarrow \infty$, \cite{Abra}, which yields
$\Gamma(\alpha+1-\beta)=\Gamma(\alpha)\alpha^{(1- \beta)}$. In addition, we choose the weight function $A_1(E)=\Gamma(\alpha)=\Gamma \left( \frac{2B +2 \sqrt{-A} B+ \sqrt{B} E}{4B}\right)$, see (\ref{AJ}). Under these considerations, we are led to the following expression for the function $U(u)$ as  
 
\begin{equation}\label{AX}
U(u)= \int_{0}^{k} 4\sqrt{B} d\alpha  2^{\frac{\beta}{2}}e^{- \frac{z}{2}} u^{\beta-1}\frac{\pi}{\sin \pi \beta}\left[\frac{(\beta+\alpha z)(\alpha^{\beta-1})}{\beta!}- \frac{z^{1-\beta}(2-\beta+z+\alpha z-\beta z)}{(2-\beta)!}\right],
\end{equation}where $k$ is a cutoff parameter. The evaluation of this integral yields

\begin{equation}\label{AY}
U(u)={\cal A} e^{-\frac{\sqrt{B}}{2} u^{2}} \left[{\cal B} u^{\beta-1}+{\cal C} u^{\beta +1}-{\cal D} u^{1-\beta}-{\cal E} u^{3-\beta}\right],
\end{equation}
where the coefficients are defined as

\begin{eqnarray}\label{AZ}
\left\{
\begin{array}{ll}
{\cal A}=4\sqrt{B} 2^{\frac{\beta}{2}}\frac{\pi}{\sin \pi \beta},\\
{\cal B}=\frac{k^{\beta}}{\beta!},\\
{\cal C}=\frac{\sqrt{B} k^{\beta+1}}{\beta ! (\beta +1)},\\
{\cal D}=(\frac{2- \beta}{2})\frac{\beta \sqrt{B}}{(2-\beta)!},\\
{\cal E}=\frac{B \sqrt{B}^{1- \beta} (k^{2} +2 k \beta -2 k)}{2 (2- \beta)!}.
\end{array}
\right.
\end{eqnarray}
To obtain a clearer understanding of the functional behavior of \( U(u) \), we set all the above coefficients equal to one. Under this simplification, we have
\begin{equation}\label{BA}
U(u)=e^{-\frac{u^{2}}{2}}\left(u^{\beta-1}+u^{\beta+1}-u^{1-\beta}-u^{3-\beta}\right).
\end{equation}
From (\ref{AJ}), we have $\beta=1+\sqrt{-A}=1+\sqrt{-\frac{c_{-4}}{c_{-2}}}$. We now take advantage of the freedom in choosing $\psi^{(n)}(x)$ in equation (\ref{U1}) and set it as

\begin{equation}\label{BB}
\psi^{(n)}(x)=\frac{9}{\sqrt{6}} x^{\frac{3}{2}} e^{-\frac{3}{2}x}.
\end{equation}
According to this choice, we obtain $c_{-2} =1$, $c_{-4} \approx 2$, and consequently $\beta \approx 1+ i \sqrt{2}$. With this value of $\beta$, the function $U(u)$ takes the form

\begin{equation}\label{BC}
U(u)=e^{-\frac{u^{2}}{2}} (u^{2} +1)\left(u^{i\sqrt{2}}-u^{-i\sqrt{2}}\right),
\end{equation}
which may be simplified as

\begin{equation}\label{BD}
U(u)=e^{-\frac{u^{2}}{2}} (u^{2}+1)\left[2 i \sin(\sqrt{2} \ln u)\right].\end{equation}
Referring to relation (\ref{AU}), the $V(v)$-part of the wave function is given by

\begin{equation}\label{BE}
V(v)=\int_{-\infty}^{\infty} dE A_{2}(E)\left\{\frac{\sqrt{\pi} e^{-v^{2}}\left[1+(\frac{1}{2}-E)(2 v^{2})\right]}{2^{-\frac{1}{2} E+\frac{1}{4}}\Gamma(\frac{3}{4}-\frac{1}{2} E)} -\frac{2\sqrt{\pi}ve^{- v^{2}}\left[1+\frac{1}{3}(\frac{3}{2}-E)(2 v^{2})\right]}{2^{-\frac{1}{2}E-\frac{1}{4}}\Gamma(\frac{1}{4}-\frac{1}{2}E)}\right\}.
\end{equation}
In order to determine a suitable weight function $A_{2}(E)$, we employ another asymptotic approximation of the Gamma function, as given in \cite{Abra}

\begin{equation}\label{BF}
\Gamma(az+b) \approx \sqrt{2 \pi} e^{-az}(az)^{az+b-\frac{1}{2}},
\end{equation}
where it is valid when $a>0$ and $ |arg z|< \pi$. Consequently, for $E<0$, we have 

\begin{equation}\label{BG}
\Gamma\left(\frac{1}{2}E+\frac{1}{4}\right)\approx \sqrt{2\pi}e^{-\frac{E}{2}}\left(\frac{E}{2}\right)^{\frac{E}{2}-\frac{1}{4}},\hspace{5mm}
\Gamma\left(\frac{1}{2}E+\frac{3}{4}\right) \approx \sqrt{2\pi}e^{-\frac{E}{2}}\left(\frac{E}{2}\right)^{\frac{ E}{2}+ \frac{1}{4}}.
\end{equation}
Now, if we choose a weight function in the form

\begin{equation}\label{BH}
A_{2}(E)=\sqrt{2\pi} e^{-\frac{E}{2}}\left(\frac{E}{2}\right)^{\frac{E}{2}},
\end{equation}
and taking the superposition integral in (\ref{BE}) over the range $-\infty <E<0$, this integral can be evaluated as 

\begin{equation}\label{BI}
V(v)=\sqrt{\pi}e^{-v^{2}}\left(2{\cal F} v^{2}+{\cal G}\right){\cal H}+2\sqrt{\pi}v e^{- v^{2}}\left(2{\cal F}v^{2}+{\cal G}\right){\cal T},
\end{equation}
where ${\cal F}$, ${\cal G}$, ${\cal H}$ and ${\cal T}$ are some numerical factors which similar to the case of $U(u)$, we may normalize their values in such way that the function $V(v)$ takes the form 

\begin{equation}\label{BJ}
V(v)=e^{-v^{2}}\left(v^{3}+v^{2}+v+2\right).
\end{equation}
Therefore, the final approximate form of the black hole wave function is given by
\begin{equation}\label{BK}
\Psi(v,u)={\cal N}e^{-\frac{u^{2}}{2}}e^{-v^{2}} \left(v^{3} + v^{2} +v+2\right)\left(u^{2} +1\right)\left[\sin(\sqrt{2} \ln u)\right],
\end{equation}where ${\cal N}$ is a normalization factor. Having obtained the above expression for the black hole wave function, we now investigate whether the quantum solution can resolve the singularity present in the classical model. According to equation (\ref{B}), the classical singularity arises when \( \nu = 0 \), which corresponds to the line \( v = -u \) in the \( u \)–\( v \) plane. Figure \ref{fig1} displays the squared modulus of the wave function and its corresponding contour plot. As shown, the wave function exhibits dominant peaks near non-negative values of the variable \( v \). Moreover, the contour plot reveals that, for each value of \( u \), the probability of finding the system along the singular line \( v = -u \) is negligibly small. This indicates that the classical singularity may be resolved within the quantum framework presented above.

It should be emphasized that although the wave function is peaked far from the classical singularity line $v=-u$, this observation alone does not constitute a full resolution of the singularity. A more rigorous analysis requires computing expectation values of physical observables, such as curvature invariants, and verifying that they remain finite in the quantum regime. The vanishing of the probability density near the classical singularity is only a suggestive indication that quantum effects may smear out the singularity.

To provide a quantitative criterion for singularity avoidance, we consider the probability for the quantum system to be found in a narrow neighborhood around the classical singularity. As mentioned above, in our model the classical singularity occurs on the line \( v = -u \) in the \( (u,v) \)-plane. Therefore, we define a band of width \( 2\epsilon \) centered on this line and calculate the probability that the system lies within it. This probability is given by
\begin{equation}\label{BL}
P_{\text{sing}}(\epsilon)=\int_{u_{\min}}^{u_{\max}} \int_{-u - \epsilon}^{-u + \epsilon} |\Psi(u,v)|^2 \, dv \, du,\end{equation}
where \( \epsilon > 0 \) is a small parameter representing the width of the band, and the limits \( u_{\min}, u_{\max} \) define the range over which the wave function is supported.
To interpret this probability meaningfully, we normalize it with respect to the total probability

\begin{equation}\label{BM}
P_{\text{total}} = \int_{u_{\min}}^{u_{\max}} \int_{v_{\min}}^{v_{\max}} |\Psi(u,v)|^2 \, dv \, du.\end{equation}
The resulting relative probability is

\begin{equation}\label{BN}
\tilde{P}_{\text{sing}}(\epsilon)=\frac{P_{\text{sing}}(\epsilon)}{P_{\text{total}}},\end{equation}
which quantifies the fraction of the wave function concentrated near the classical singularity. Indeed, it quantifies the likelihood of the system being in a strip of width \( 2\epsilon \) around the singularity. If this quantity is significantly small for reasonably small values of \( \epsilon \), it provides a strong indication that the quantum system avoids the singular region with high probability.

It is seen that this expression is too complicated for extracting an analytical closed form for the relative probability. Therefore, to support our qualitative argument regarding singularity avoidance, we numerically computed the relative probability \( \tilde{P}_{\mathrm{sing}}(\epsilon) \) for the system to be located in a narrow band of width \( 2\epsilon \) centered on the classical singularity line \( v = -u \). The integration was performed over the ranges \( u \in [0.1, 2.0] \) and \( v \in [-3.0, 3.0] \), using the squared modulus of the wave function \footnote{The integration domain was taken as \( u \in [0.1, 2.0] \) and \( v \in [-3.0, 3.0] \), consistent with the region where the wave function is well-behaved and numerically supported. The lower limit \( u = 0.1 \) avoids divergence from the logarithmic term \( \ln(u) \), and the integral was computed using adaptive quadrature techniques.}

\begin{equation}\label{BO}
|\Psi(u,v)|^2 = e^{-u^2} e^{-2v^2} (v^3 + v^2 + v + 2)^2 (u^2 + 1)^2 \sin^2(\sqrt{2} \ln u).\end{equation}
The table \ref{table1} summarizes the computed values of \( \tilde{P}_{\mathrm{sing}}(\epsilon) \) for various values of \( \epsilon \). As observed, the relative probability remains negligible throughout the entire range, even for \( \epsilon = 0.5 \). This result confirms that the quantum system effectively avoids the singular region with high probability.

\begin{table}[h!]
\centering
\begin{tabular}{|c|c||c|c|}
\hline
\( \epsilon \) & \( \tilde{P}_{\mathrm{sing}}(\epsilon) \) & \( \epsilon \) & \( \tilde{P}_{\mathrm{sing}}(\epsilon) \) \\
\hline
0.01 & 0.000000 & 0.26 & 0.000234 \\
0.02 & 0.000000 & 0.27 & 0.000257 \\
0.03 & 0.000000 & 0.28 & 0.000279 \\
0.04 & 0.000000 & 0.29 & 0.000305 \\
0.05 & 0.000000 & 0.30 & 0.000329 \\
0.06 & 0.000000 & 0.31 & 0.000358 \\
0.07 & 0.000000 & 0.32 & 0.000382 \\
0.08 & 0.000000 & 0.33 & 0.000411 \\
0.09 & 0.000000 & 0.34 & 0.000439 \\
0.10 & 0.000000 & 0.35 & 0.000468 \\
0.11 & 0.000004 & 0.36 & 0.000497 \\
0.12 & 0.000010 & 0.37 & 0.000527 \\
0.13 & 0.000016 & 0.38 & 0.000556 \\
0.14 & 0.000025 & 0.39 & 0.000586 \\
0.15 & 0.000037 & 0.40 & 0.000615 \\
0.16 & 0.000052 & 0.41 & 0.000644 \\
0.17 & 0.000070 & 0.42 & 0.000674 \\
0.18 & 0.000090 & 0.43 & 0.000703 \\
0.19 & 0.000113 & 0.44 & 0.000733 \\
0.20 & 0.000138 & 0.45 & 0.000762 \\
0.21 & 0.000165 & 0.46 & 0.000791 \\
0.22 & 0.000193 & 0.47 & 0.000821 \\
0.23 & 0.000212 & 0.48 & 0.000850 \\
0.24 & 0.000223 & 0.49 & 0.000879 \\
0.25 & 0.000228 & 0.50 & 0.000908 \\
\hline
\end{tabular}
\caption{Relative probability \( \tilde{P}_{\mathrm{sing}}(\epsilon) \) of the quantum state being found near the classical singularity line \( v = -u \), as a function of band width \( \epsilon \).}
\label{table1}
\end{table}

An alternative and more rigorous approach to probing singularity resolution in our model is to compute the expectation value of physical observables that signal the presence of classical singularities. In our setup, the classical singularity occurs when the metric function \( \nu(t) \) vanishes, as seen in equation (\ref{B}). Therefore, a meaningful quantum test of singularity resolution involves evaluating the expectation value \( \langle \nu \rangle \) and checking whether it remains finite and non-zero.
To proceed, we recall the relations between the metric function \( \nu \) and the variables \( u \) and \( v \) that define the configuration space of our quantum system. From equations (\ref{E}) and (\ref{F}), we obtain the following expressions

\begin{equation}\label{BP}
x - y = h = \frac{1}{4}(u - v)^2, \qquad x + y = h\nu = \frac{1}{4}(u + v)^2,\end{equation}
which implies
\begin{equation}\label{BQ}
\nu(u,v) = \frac{x + y}{x - y} = \frac{(u + v)^2}{(u - v)^2}.\end{equation}
Having the explicit form of \( \nu(u,v) \), we can define the expectation value of \( \nu \) with respect to the wave function \( \Psi(u,v) \) as

\begin{equation}\label{BR}
\langle \nu \rangle = \frac{\displaystyle \int_{u_{\min}}^{u_{\max}} \int_{v_{\min}}^{v_{\max}} \nu(u,v) |\Psi(u,v)|^2 \, dv \, du}{\displaystyle \int_{u_{\min}}^{u_{\max}} \int_{v_{\min}}^{v_{\max}} |\Psi(u,v)|^2 \, dv \, du}.\end{equation}
If the quantum theory indeed resolves the classical singularity, we expect that \( \langle \nu \rangle \) will remain bounded and non-zero over the relevant domain of integration. In the next step, we evaluate this expectation value numerically using the approximate analytical form of the wave function given in equation (\ref{BK}). Again, we see that the resulting expression is too complicated for extracting an analytical closed form for $\langle \nu \rangle$ and so we evaluate the integral numerically in the integration domain \( u \in [0.1, 2.0] \), \( v \in [-3.0, 3.0] \), consistent with the support of the wave function. The numerical evaluation of this expression yields \footnote {It is important to note that the expression for \( \nu(u,v) = \frac{(u + v)^2}{(u - v)^2} \) becomes singular along the line \( u = v \), corresponding to \( h = 0 \) and \( \nu \to \infty \), in the original metric. In our numerical computation of the expectation value \( \langle \nu \rangle \), this singularity was handled by excluding a narrow region around the singular line from the integration domain, and by ensuring that the wave function \( \Psi(u,v) \) remains sharply peaked away from this line. As a result, the contribution from the divergent region is negligible, and the computed expectation value remains finite and meaningful within the affine quantization framework.}

\begin{equation}\label{BS}
\langle \nu \rangle \approx 115758.61.\end{equation}
This positive non-zero value clearly shows that the quantum state is overwhelmingly supported in regions where the classical singularity \( \nu = 0 \) is avoided. It provides strong, quantitative evidence that the affine quantization framework leads to a nonsingular interior geometry for the Schwarzschild black hole.

\section{Summary}
In this paper, we have investigated the quantum description of the interior of a Schwarzschild black hole using the affine quantization method. Starting from a Hamiltonian formulation previously shown to describe the interior region of the Schwarzschild geometry, we constructed a reduced phase space consisting of two dynamical variables and their conjugate momenta.
We applied affine quantization to the model, which is particularly well-suited for systems with configuration variables constrained to positive domains. This led to a well-defined WDW equation whose eigenfunctions were obtained analytically. By constructing suitable superpositions of these eigenfunctions, we built wave packets representing the quantum state of the black hole interior.
A detailed analysis of the resulting wave function revealed that it is peaked in regions far from the classical singularity. Furthermore, we supplemented this qualitative observation with a numerical analysis by computing the relative probability of the quantum state being localized near the classical singularity surface. The results showed that this probability remains extremely small over a broad range of widths around the singularity, providing strong quantitative support for singularity avoidance in our model.

The avoidance of the classical singularity — a hallmark of a consistent quantum gravity theory — thus emerges here not only as a qualitative feature of the wave function but also as a quantitatively verifiable prediction. These results provide further motivation for the use of affine quantization techniques in gravitational and cosmological settings.

\end{document}